\documentclass[aps, prl, twocolumn, superscriptaddress, 10pt, floatfix]{revtex4-1}
\usepackage{graphicx, color}
\usepackage{natbib}
\usepackage{amsmath, amssymb}
\usepackage[colorlinks = true,
			citecolor = blue,
			linkcolor = blue]{hyperref}

\newcommand{\refeq}[1]{eq.~(\ref{#1})}

\newcommand{\reffig}[1]{Fig.~\ref{#1}}
\newcommand{\refFig}[1]{Figure~\ref{#1}}

\newcommand{\Is}{I_{\rm sat}}
\newcommand{\Id}{I_{\rm ob}}
\newcommand{\Ed}{E_{\rm ob}}
\newcommand{\kd}{k_{\rm ob}}
\newcommand{\If}{I_{\rm f}}
\newcommand{\Ef}{E_{\rm f}}
\newcommand{\kf}{k_{\rm f}}
\newcommand{\cs}{c_{\rm s}}

\newcommand{\nr}{n_{\rm e}}
\newcommand{\ttt}[1]{\times 10^{#1}}

\begin{document}
\title{Superfluid motion and drag-force cancellation in a fluid of light}
\author{Claire Michel}
\affiliation{Universit\'e C\^ote d'Azur, CNRS, Institut de Physique de Nice, France}
\author{Omar Boughdad}
\affiliation{Universit\'e C\^ote d'Azur, CNRS, Institut de Physique de Nice, France}
\author{Mathias Albert}
\affiliation{Universit\'e C\^ote d'Azur, CNRS, Institut de Physique de Nice, France}
\author{Pierre-\'Elie Larr\'e}
\affiliation{Laboratoire Kastler-Brossel, Sorbonne Universit\'es, CNRS, ENS-PSL Research University, Coll\`ege de France, 4 Place Jussieu, 75005 Paris, France}
\author{Matthieu Bellec}
\affiliation{Universit\'e C\^ote d'Azur, CNRS, Institut de Physique de Nice, France}

\begin{abstract}
	Quantum fluids of light merge many-body physics and nonlinear optics, through the study of light propagation in a nonlinear medium under the shine of quantum hydrodynamics. One of the most outstanding evidence of light behaving as an interacting fluid is its ability to carry itself as a superfluid. Here, we report a direct experimental detection of the transition to superfluidity in the flow of a fluid of light past an obstacle in a bulk nonlinear crystal. In this cavityless all-optical system, we extract a direct optical analog of the drag force exerted by the fluid of light and measure the associated displacement of the obstacle. Both quantities drop to zero in the superfluid regime characterized by a suppression of long-range radiation from the obstacle. The experimental capability to shape both the flow and the potential landscape paves the way for simulation of quantum transport in complex systems.
\end{abstract}
\maketitle

Superfluidity was originally discovered in 1938~\cite{Kapitza1938} when a \textsuperscript{4}He fluid cooled under its $\lambda$-point flowed in a nonclassical way along a capillary~\cite{Allen1938}. This was the trigger for the development of many experiments genuinely realized with quantum matter, as with \textsuperscript{3}He fluids~\cite{Osheroff1972} or ultracold atomic vapors~\cite{Pitaevskii2016, Bloch2012}. The superfluid behavior of mixed light-matter cavity gases of exciton-polaritons was also extensively studied~\cite{Amo2009, Amo2011}, leading to the emergent field of ``quantum fluids of light''~\cite{Carusotto2013}. 
Before being theoretically developed for cavity lasers \cite{Vaupel1996, Chiao1999}, the idea of a superfluid motion of light originates from pioneering studies in cavityless all-optical configurations \cite{Pomeau1993} in which the hydrodynamic nucleation of quantized vortices past an obstacle when a laser beam propagates in a bulk nonlinear medium was investigated \cite{Frisch1992}. In such a cavityless geometry, the paraxial propagation of a monochromatic optical field in a nonlinear medium may be mapped onto a two-dimensional Gross-Pitaevskii-type evolution of a quantum fluid of interacting photons in the plane transverse to the propagation \cite{Pitaevskii2016}. The intensity, the gradient of the phase and the propagation constant of the optical field assume respectively the roles of the density, the velocity and the mass of the quantum fluid, and the photon-photon interactions are mediated by the optical nonlinearity. It took almost twenty years for this idea to spring up again~\cite{Wan2007, Khamis2008, Leboeuf2010, Carusotto2014}, driven by the emergence of advanced laser-beam-shaping technologies allowing to precisely tailor both the shape of the flow and the potential landscape. 
\begin{figure}[b]
	\centering
	\includegraphics{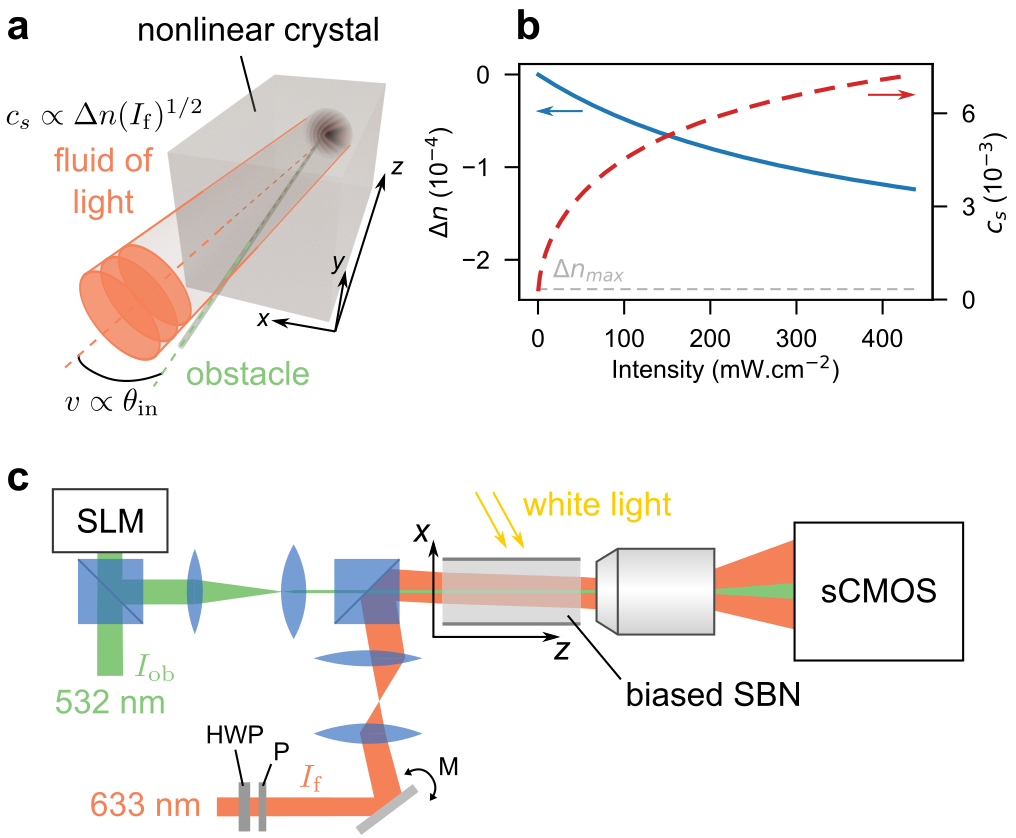}
	\caption{\textbf{Realization of a fluid of light in a propagating geometry and nonlinear response of the bulk crystal.}
		(\textbf{a}) Sketch of the fluid of light (red beam) flowing past an obstacle (green beam). The input velocity $v$ of the fluid of light is proportional to the input angle $\theta_{\rm in}$. The sound velocity $\cs$ depends on the intensity $\If$ of the red beam. 
		(\textbf{b}) \textit{Blue curve}. Calculated optical-index variation $\Delta n$ with respect to a laser intensity $I$ for the nonlinear photorefractive response of the medium. \textit{Red dashed curve}. Corresponding sound velocity $\cs$. 
		(\textbf{c}) Experimental setup. The green beam is shaped by the spatial light modulator (SLM) to create a $z$-invariant optical defect acting as a localized obstacle in the transverse plane. The red beam is a large gaussian beam and creates the fluid of light. $\If$ is controlled by a half-waveplate (HWP) and a polarizer (P). $\theta_{\rm in}$ is tuned by rotating a mirror (M) imaged at the input of the crystal via a telescope. Both are propagating simultaneously through a biased SBN photorefractive crystal and imaged on a sCMOS camera. The white light controls the saturation intensity of the crystal.
	}
	\label{fig:sketch}
\end{figure}

The ways of tracking light superfluidity are manifold. Recently, superfluid hydrodynamics of a fluid of light has been studied in a nonlocal nonlinear liquid through the measurement of the dispersion relation of its elementary excitations~\cite{Vocke2015} and the detection of a vortex nucleation in the wake of an obstacle~\cite{Vocke2016}. The stimulated emission of dispersive shock waves in nonlinear optics was also studied in the context of light superfluidity~\cite{Wan2007}. 
However, one of the most striking manifestations of superfluidity --- which is the ability of a fluid to move without friction~\cite{Leggett1999} --- has never been directly observed in a cavityless nonlinear-optics platform. 
A direct consequence of this feature is the absence of long-range radiation in a slow fluid flow past a localized obstacle. In optical terms, this corresponds to the absence of light diffraction from a local modification of the underlying refractive index in the plane transverse to the propagation. On the contrary, in the ``frictional'', nonsuperfluid regime, light becomes sensitive to such an index modification and diffracts while hitting it.

\section*{Results.}
\noindent \textbf{Hydrodynamics of light.}
Here, we make use of a biased photorefractive crystal which is, thanks to its controllable nonlinear optical response, convenient for probing the hydrodynamic behavior of light~\cite{Wan2007, Sun2012}. 
As sketched in \reffig{fig:sketch}a and detailed in \reffig{fig:sketch}c, a local drop of the optical index is photo-induced by a narrow beam in the crystal and creates the obstacle. Simultaneously, a second, larger monochromatic beam is sent into the crystal and creates the fluid of light.
We report a direct observation of a superfluid regime characterized by the absence of long-range radiation from the obstacle. This regime is usually associated to the cancellation of the drag force experienced by the obstacle, as studied for \textsuperscript{4}He~\cite{Allum1977}, ultracold atomic gases~\cite{Raman1999, Pavloff2002, Miller2007, Engels2007, Desbuquois2012}, or cavity exciton-polaritons~\cite{Wouters2010, Berceanu2012, VanRegemortel2014, Larre2012}. 
In our cavityless all-optical system, we extract on the one hand a quantity corresponding to the optical analog of this force and measure on the other hand the associated obstacle displacement. 
For the first time, at least within the framework of fluids of light, we observe that this displacement is nonzero in the nonsuperfluid case and tends to vanish while reaching the superfluid regime. 

The propagation of the fluid-of-light beam in the paraxial approximation is ruled by a two-dimensional Gross-Pitaevskii-type equation (also known as a nonlinear Schr\"odinger-type equation):
\begin{equation}
	i\partial_z \Ef = -\frac{1}{2\nr\kf}\nabla^2 \Ef - \kf\Delta n(\Id)\Ef - \kf\Delta n(\If)\Ef
	\label{eq:schro}
\end{equation}
The propagation coordinate $z$ plays the role of time. The transverse-plane coordinates $\mathbf{r}=(x,y)$ span the two-dimensional space in which the fluid of light evolves. The propagation constant $\nr\,\kf=\nr \times 2\pi/\lambda_{\rm f}$ of the fluid-of-light beam propagating in the crystal of refractive index $\nr$ is equivalent to a mass; the associated Laplacian term describes light diffraction in the transverse plane. 
The density of the fluid is given by the intensity $\If\propto|\Ef|^{2}$. Its velocity corresponds to the gradient of the phase of the optical field. At the input, it is simply given by $v\simeq\theta_{\rm in}/\nr$, with $\theta_{\rm in}$ the angle between the fluid-of-light beam and the $z$ axis (see Supplementary Sec.~S1 for more details).
The local refractive index depletion $\Delta n[\Id(\mathbf{r})]<0$ is induced by the obstacle beam of intensity $\Id(\mathbf{r})$. The self-defocusing nonlinear contribution $\Delta n(\If)<0$ to the total refractive index provides repulsive photon-photon interactions and ensures robustness against modulational instabilities~\cite{Larre2017}. 
From the latter, we define an analog healing length $\xi=[\nr\,\kf\times\kf\,|\Delta n(\If)|]^{-1/2}$, which corresponds to the smallest length scale for intensity modulations, and an analog sound velocity $\cs=(\nr\,\kf\times\xi)^{-1}=[|\Delta n(\If)|/\nr]^{1/2}$ for the fluid of light~\cite{Pitaevskii2016, Carusotto2014}. 
The photorefractive nonlinear response of the material, $\Delta n(I)$, is plotted in blue in \reffig{fig:sketch}b as a function of the laser intensity $I$ (see the Methods section for details). In the same figure, the red dashed curve represents the speed of sound $\cs(I)$.


When the obstacle is infinitely weakly perturbing, Landau's criterion for superfluidity~\cite{Leggett1999} applies and the so-called Mach number $v/c_{\mathrm{s}}$ mediates the transition around $v/\cs=1$ from a nonsuperfluid regime at large $v/c_{\mathrm{s}}$ to a superfluid regime at low $v/c_{\mathrm{s}}$. 
Generally this condition is not fulfilled and the actual critical velocity is lower than the sound velocity $\cs$.~\cite{Feynman1955,Pitaevskii2016}. 
This is the case in the present work for two main reasons. First, we consider a weakly but finite perturbing obstacle. It means a small variation of the refractive index $\Delta n\left[\Id(\mathbf{r})\right] = -2.2 \times 10^{-4}$ and a radius of $6$ $\mu$m comparable to $\xi$ (see Methods and Supplementary Sec.~S2). 
Note however that the perturbation is weak enough for the transition not to be blurred by the emission of nonlinear excitations like vortices or solitons.
Second, remaining within Landau's picture, the speed of sound is here defined for $\If$ measured at its maximum value, at $z=0$, whereas the latter naturally suffers from linear absorption and self-defocusing along the $z$ axis.
\begin{figure*}
	\centering
	\includegraphics{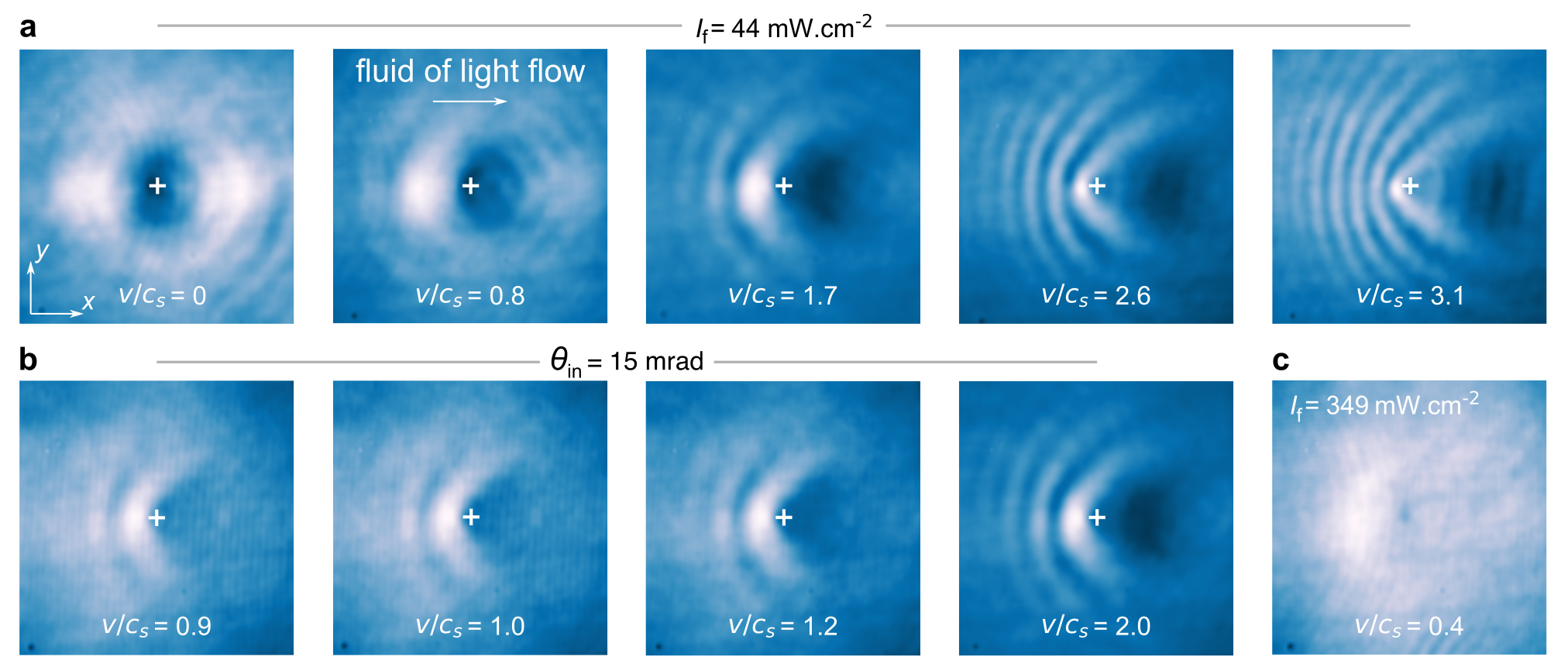}
	\caption{\textbf{Spatial distribution of the output intensity of the fluid of light for various input conditions.}
	The fluid of light flows from left to right. The white crosses at the center of the images indicate the position of the obstacle. Each image is 330 $\times$ 330 $\mu$m$^2$.
		(\textbf{a}) At a fixed input intensity $\If$, the input angle $\theta_{\rm in}$ of the beam creating the fluid of light is tuned to vary the Mach number $v/\cs$ from 0 to 3.1.
		(\textbf{b}) Similarly, at a fixed input angle $\theta_{\rm in}$, $\If$ is progressively decreased to change $v/\cs$ from 0.9 to 2.0.
		(\textbf{c}) For large $\If$, the fluid of light is clearly in the superfluid regime at $v/\cs = 0.4$. The remaining lack of uniformity upstream from the obstacle is attributed to propagation losses due to linear absorption.
	}
	\label{fig:images}
\end{figure*}

\noindent \textbf{Probing the transition to superfluidity.}
The ratio $v/\cs$ is experimentally controlled by the incidence angle $\theta_{\rm in}$ and by the input intensity $\If$ of the fluid-of-light beam. \refFig{fig:images} presents typical experimental results for the spatial distribution of the light intensity observed at the output of the crystal for various input conditions. 
\refFig{fig:images}a displays the output spatial distributions of intensity for different fluid velocities $v$ at a fixed speed of sound, $\cs = 3.2 \ttt{-3}$. This allows to vary $v/\cs$ from 0 to 3.1. As $v$ increases, diffraction appears in the transverse plane, and progressively manifests as a characteristic cone of fringes upstream from the obstacle~\cite{Carusotto2006, Khamis2008, Carusotto2014}.
Another way to probe the transition is to fix the transverse velocity $v$ and to vary the sound velocity $\cs$ through the variation of the intensity of the fluid-of-light beam. Although the two ways of varying $v/\cs$ are not equivalent, as we shall discuss later, the results shown in \reffig{fig:images}b are similar with the interference pattern becoming more and more pronounced as $v/\cs$ increases.  
\refFig{fig:images}c represents the intensity distribution at the output of the crystal for $v/\cs=0.4$. Long-range radiation upstream from the obstacle is no longer present in this case, indicating a superfluid motion of light. The lack of uniformity of the intensity upstream from the obstacle is due to the intrinsic linear absorption of the material \cite{Larre2012}.

\noindent \textbf{Drag-force and obstacle displacement.}
In the supersonic regime, the intensity modulation of the fluid of light flowing around the obstacle induces a local optical-index modification of the material. This modification influences the propagation of the beam responsible for the obstacle, for which a transverse displacement is expected. On the contrary, in the superfluid regime, the absence of long-range intensity perturbations implies no local variation of the optical index and then one does not await for any displacement of the obstacle beam.
 
As theoretically investigated in~\cite{Larre2015} for a material obstacle (here, we rather consider an all-optical obstacle), the local intensity difference for the fluid of light between the front ($I_+$) and the back ($I_-$) of the obstacle, $I_+ - I_-$, is proportional to the dielectric force experienced by the obstacle. This force turns out to be closely analogous to the drag force that an atomic Bose-Einstein condensate exerts onto some obstacle.
%
\begin{figure}
	\centering
	\includegraphics{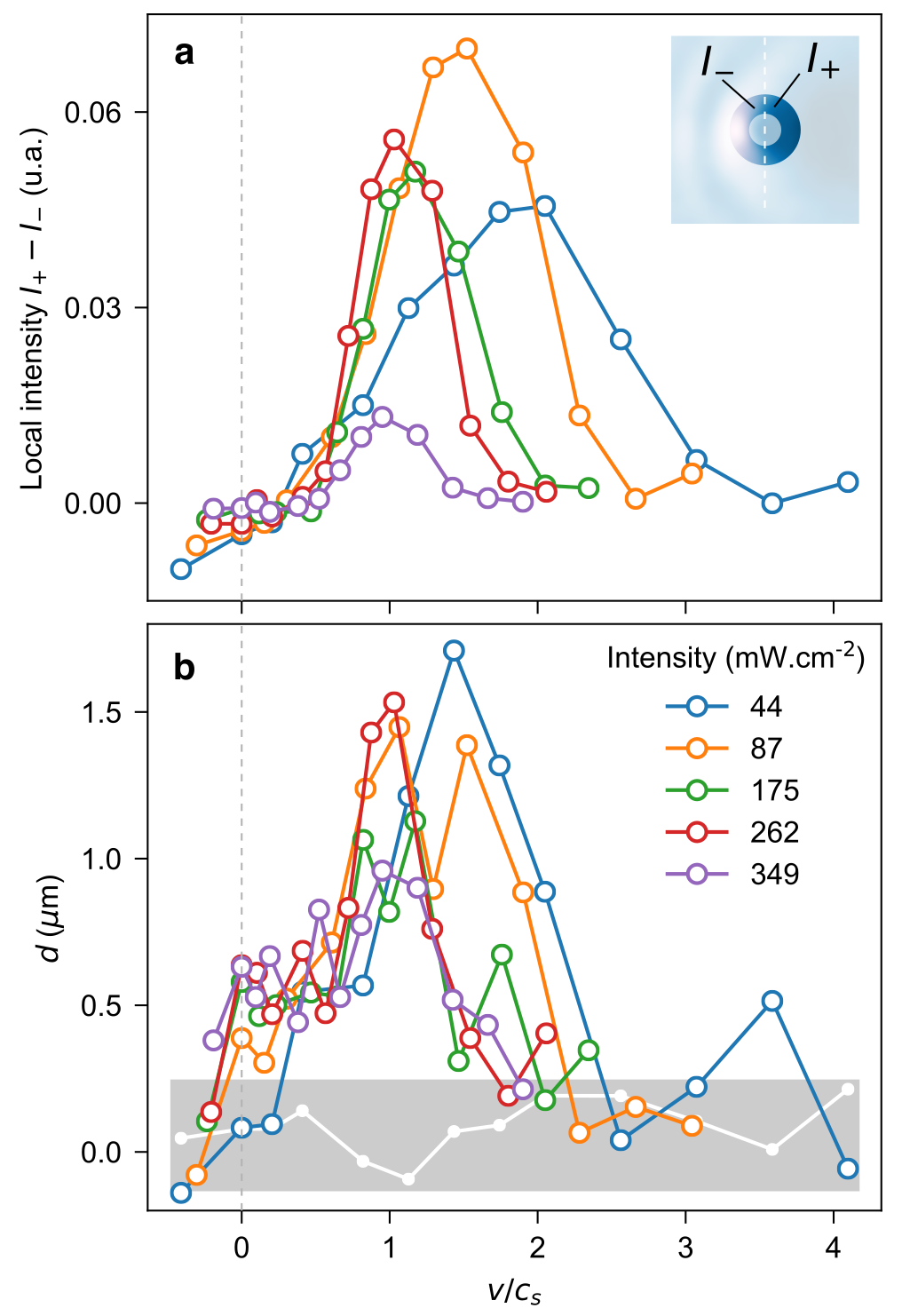}
	\caption{\textbf{Optical analog of the drag force exerted by the fluid and associated displacement of the obstacle.} 
		(\textbf{a}) Local intensity difference $I_+ - I_-$ extracted from the experimental images of the intensity of the fluid-of-light beam measured at the crystal's output for various input conditions ($\If$ ranging from 44 to 349 mW.cm$^{-2}$ and $v/\cs$ ranging from -0.41 to 4.10). Inset: the original image is cropped around the optical defect and integrated over two regions, downstream ($I_-$) and upstream ($I_+$). The typical integration area is of the order of $\xi$. The gray dotted line corresponds to $v/\cs = 0$. 
		(\textbf{b}) Measurement of the transverse displacement of the obstacle induced by the local modulation of the intensity of the fluid of light for various input conditions. The gray box defines the typical uncertainty in the measured quantities, the white points corresponding to the displacement along the $y$ axis for $\If = 44$ mW.cm$^{-2}$, which is expected to be zero.}
	\label{fig:displacement_local_int}
\end{figure}
\refFig{fig:displacement_local_int}a depicts the variation of $I_+ - I_-$, measured at the output of the crystal, as a function of $v/\cs$ for various initial conditions. As illustrated in the inset, both intensities are integrated over a typical distance of the order of $\xi$ surrounding the obstacle.  
For all intensities, we observe a rather smooth, but net transition for $v$ slightly smaller than $\cs$. 
The increasing tendency for low Mach numbers is associated to linear absorption, as discussed in the context of cavity quantum fluids of light~\cite{Wouters2010, Larre2012, Berceanu2012}.
The well-known decreasing tendency at large Mach numbers is also observed. Indeed, the obstacle can always be treated as a perturbation at large velocities and the associated drag force resultingly decreases \cite{albert2008}.
As the intensities increase, one can see that the local intensity difference sticks to zero for non-zero values of $v/\cs$, as predicted for the drag fore in a superfluid regime.
Moreover, \reffig{fig:displacement_local_int}a shows that the curves with different intensities $\If$, although renormalized by the respective sound velocity $c_s$, do not fall on a single universal curve. This is due to the fact that changing the intensity also affects crucial quantities like the healing length $\xi$ and the relative strength of the obstacle with respect to the nonlinear term, $\Delta n(\Id) / \Delta n(\If)$.
While the drop of this force is among the main signatures of superfluidity in material fluids, so far this is the first experiment on fluids of light investigating it.

To go one step further, we probe the corresponding transverse displacement of the obstacle, independently on the measurement of $I_+-I_-$.
By assuming that the transverse component of the fluid-of-light beam is non-zero only along the $x$ axis, we denote by $\langle x \rangle = \int x |\Ed|^2 \, dx$ the position of the centroid of the obstacle beam. Using an optical equivalent of the Ehrenfest relations, one can derive the following equation of motion (see Supplementary Sec.~S3 for full derivation):
\begin{equation}
\nr\,\partial_{zz}\langle x \rangle = \partial_x [\Delta n(\If)].
\label{eq:motion}
\end{equation}
This means that the all-optical obstacle is sensitive to the surrounding refractive index potential resulting from the spatial distribution of intensity of the beam creating the fluid of light and might move of a distance $d=\langle x\rangle-x_0$ from its initial position $x_0$ in the transverse plane.
The measurement of $d$ for various conditions in the case of an obstacle evolving in a fluid of light at rest allows to validate such an experimental approach and to extract experimental parameters as $\Is$ and $\Delta n_{\rm max}$ (see Methods and Supplementary Sec. S3).

Figure~\ref{fig:displacement_local_int}b shows the transverse displacement measured in a moving fluid of light varying the Mach number $v/\cs$ for different initial conditions. To take into account the gaussian shape of $\If$, we subtract, for each data point, the displacement measured when the influence of the obstacle on the fluid of light is negligible (i.e. very low $\Id$), as illustrated in Supplementary Sec.~S3. 
The displacement along the $y$ direction, measured for $\If = 44$ mW.cm$^{-2}$ and which is expected to be zero, is represented by the white data and makes it possible to define the typical measurement uncertainty for this experiment (gray box). The fluctuation can be attributed to the inherent imperfections of the fluid-of-light beam. 
We observe that the transverse displacement of the obstacle behaves very similarly to the intensity difference $I_+-I_-$ displayed in \reffig{fig:displacement_local_int}a. That is, an increasing displacement from almost zero in the deeply subsonic regime to maximum signal, and then a decreasing tendency in the supersonic regime. We also measured an opposite transverse displacement for negative $v/\cs$. The fact that the displacement is not purely zero in the superfluid regime is likely due to the displacement acquired during the non-stationary regime at early stage of the propagation (see Supplementary Sec.~S4 for qualitative discussion supported by numerical simulations). 
This is, to the best of our knowledge, the first observation of the displacement of an all-optical obstacle in a fluid of light.

To conclude, we reported a direct experimental observation of the transition from a ``frictional'' to a superfluid regime in a cavityless all-optical propagating geometry. We performed a quantitative study by extracting an optical equivalent of the drag force that the fluid of light exerts on the obstacle. This result is in very good agreement with an independent measurement that consists in studying the transverse displacement of the obstacle surrounded by the fluid of light. 
We restricted the present study to the case of a weakly perturbing obstacle but our experimental setup allows to reach the turbulent regime associated to vortex generation through the induction of a greater optical-index depletion. 
On the other hand, a different shaping of the beam creating the obstacle will allow to generate any kind of optical potential and to extend the study to imaging through disordered environments.

\section*{Methods}
\label{sec:methods}
\noindent \textbf{Experimental setup.}
The nonlinear medium consists in a $5\times5\times10$ mm$^{3}$ strontium barium niobate (SBN:61) photorefractive crystal additionally doped with cerium (0.01\%) to enhance its photoconductivity~\cite{Buse1997} albeit it induces linear absorption (3.2 dB/cm). 
The basic mechanism of the photorefractive effect remains in the photogeneration and displacement of mobile charge carriers driven by an external electric field $E_0$. The induced permanent space-charge electric field thus implies a modulation of the refractive index of the crystal~\cite{Denz2003}, $\Delta n(I, \mathbf{r}) = -0.5\nr^3 r_{33} E_0/\left[1+I(\mathbf{r})/\Is\right]$, where $\nr$ is the optical refractive index and $r_{33}$ the electro-optic coefficient of the material along the extraordinary axis, $I(\mathbf{r})$ is the intensity of the optical beam in the transverse plane $\mathbf{r}(x,y)$, and $\Is$ is the saturation intensity which can be adjusted with a white light illumination of the crystal. 
The blue curve in \reffig{fig:sketch}b shows the saturable nonlinear response of the material $\Delta n(I)$ against the laser intensity $I$. The red dashed curve represents the sound velocity $\cs(I)$ for the saturable nonlinear response of the material $\Delta n(I)$.  The maximum value of the optical index variation is theoretically $\Delta n_{\rm max} = -2.32 \times 10^{-4}$ for $E_0 = 1.5$ kV.cm$^{-1}$.
\vspace{0.5em}

\noindent \textbf{Shaping the fluid of light and obstacle beams.}
Making use of a spatial light modulator, we produce a diffraction-free Bessel beam ($\lambda_{\rm ob}=532$ nm, $\Id = 7.6$ W.cm$^{-2}$ $\gg \Is$, green path in \reffig{fig:sketch}c). The latter creates the obstacle with a radius of 6 $\mu$m (comparable to $\xi = 6.2$ $\mu$m obtained for $\If = 349$ mW.cm$^{-2}$) that is constant all along the crystal and aligned with the $z$-direction. From \reffig{fig:sketch}b, the propagation of the obstacle beam into the crystal induces a local drop $\Delta n(\Id)=-2.2\ttt{-4}$ in the refractive index. 
A second laser ($\lambda_{\rm f}=633$ nm, red path in \reffig{fig:sketch}c) delivers a gaussian beam whose radius is extended to $270$ $\mu$m and which corresponds to the fluid-of-light beam. 
Both laser beams are linearly-polarized along the extraordinary axis to maximize the photorefractive effect.
We vary the flow velocity $v$ by changing the input angle $\theta_{\rm in}$ of the fluid-of-light beam with respect to the propagation axis $z$ (see \reffig{fig:sketch}a). 
The accessible range, tuned by rotating a mirror imaged at the input of the crystal via a telescope, goes from $\theta_{\rm in}=0$ to $\pm 23$ mrad, corresponding to $v$ ranging from $v=0$ to $v = \pm 1.3\ttt{-2}$. The sound velocity $\cs$ is controlled by the input intensity of the beam which can be tuned from $\If=0$ to $350$ mW.cm$^{-2}$ via a half-waveplate and a polarizer. The maximum value for $\cs$ is $6.8\ttt{-3}$, as plotted in Fig.~\ref{fig:sketch}b.
For the detection part, a $\times$20 microscope objective and a sCMOS camera allow to get the spatial distribution of the near-field intensity of the beams at the output of the crystal.
\vspace{0.5em}

\noindent \textbf{Displacement of the obstacle in a fluid of light at rest.}
In order to validate our experimental approach, we consider the linear propagation of the green beam creating the obstacle in the optical potential $\Delta n (\If)$ photo-induced by the fluid-of-light beam at rest ($\theta_{\rm in}=0$). In the paraxial approximation, the propagation equation reads
\begin{equation}
	i\partial_z \Ed = -\frac{1}{2\nr\kd}\nabla^2 \Ed - \kd \Delta n(\If) \Ed\;,
	\label{eq:schro_defect}
\end{equation}
with notations similar to the ones used in \refeq{eq:schro}. By assuming that the transverse component of the fluid-of-light beam is non-zero only along the $x$ axis, we denote by $\langle x \rangle = \int x |\Ed|^2 \, dx$ the position of the centroid of the obstacle beam. Using an optical equivalent of the Ehrenfest relations (see Supplementary Sec.~S3 for full derivation), one can derive from \refeq{eq:schro_defect} the following equation of motion: $(\nr\,\kd)\,\partial_{zz}\langle x \rangle = -\partial_x [-\kd\,\Delta n(\If)]$. Assuming that that $\Delta n$ is $z$-independent, which is valid in the here-considered linear propagation of the obstacle beam, we readily obtain
\begin{equation}
	d=\langle x(z) \rangle - x_0 = \frac{1}{2}\,[\partial_x \Delta n(\If)/\nr]\,z^2
	\label{eq:displacement}
\end{equation}
where $x_0$ is the initial position of the obstacle. 
This displacement is interpreted as the consequence of a force deriving from the optical potential $-\kd\,\Delta n(\If)$, and acting on the obstacle.

\noindent The experimental measurement of $d$, for various intensities $\If$ and positions $x_0$, is presented in Fig.~\ref{fig_supp:displacement}. The experimental data are fitted, using the above expression, the saturation intensity and the maximum refractive index modification being the fitting parameters. We extract $\Is = 380 \pm 50$ mW.cm$^{-2}$ and $\Delta n_{\rm max} = 2.5 \pm 0.4 \ttt{-4}$. 
It is worth mentioning that the value of $\Is$ is used for the calculation of $\Delta n(I) $ and its deriving quantities (i.e., $\cs$ and $\xi$).

\bibliographystyle{naturemag}
\bibliography{library}

\begin{thebibliography}{10}
\expandafter\ifx\csname url\endcsname\relax
  \def\url#1{\texttt{#1}}\fi
\expandafter\ifx\csname urlprefix\endcsname\relax\def\urlprefix{URL }\fi
\providecommand{\bibinfo}[2]{#2}
\providecommand{\eprint}[2][]{\url{#2}}

\bibitem{Kapitza1938}
\bibinfo{author}{Kapitza, P.}
\newblock \bibinfo{title}{Viscosity of liquid helium below the
  $\lambda$-point}.
\newblock \emph{\bibinfo{journal}{Nature}} \textbf{\bibinfo{volume}{141}},
  \bibinfo{pages}{74} (\bibinfo{year}{1938}).

\bibitem{Allen1938}
\bibinfo{author}{Allen, J.~F.} \& \bibinfo{author}{Misener, A.~D.}
\newblock \bibinfo{title}{Flow of liquid helium ii}.
\newblock \emph{\bibinfo{journal}{Nature}} \textbf{\bibinfo{volume}{141}},
  \bibinfo{pages}{75} (\bibinfo{year}{1938}).

\bibitem{Osheroff1972}
\bibinfo{author}{Osheroff, D.~D.}, \bibinfo{author}{Richardson, R.~C.} \&
  \bibinfo{author}{Lee, D.~M.}
\newblock \bibinfo{title}{Evidence for a new phase of solid he${}^3$}.
\newblock \emph{\bibinfo{journal}{Phys. Rev. Lett.}}
  \textbf{\bibinfo{volume}{28}}, \bibinfo{pages}{885} (\bibinfo{year}{1972}).

\bibitem{Pitaevskii2016}
\bibinfo{author}{Pitaevskii, L.} \& \bibinfo{author}{Stringari, S.}
\newblock \emph{\bibinfo{title}{Bose-Einstein Condensation and Superfluidity}}
  (\bibinfo{publisher}{Oxford University Press}, \bibinfo{year}{2016}).

\bibitem{Bloch2012}
\bibinfo{author}{Bloch, I.}, \bibinfo{author}{Dalibard, J.} \&
  \bibinfo{author}{Nascimb{\`{e}}ne, S.}
\newblock \bibinfo{title}{Quantum simulations with ultracold quantum gases}.
\newblock \emph{\bibinfo{journal}{Nat. Phys.}} \textbf{\bibinfo{volume}{8}},
  \bibinfo{pages}{267} (\bibinfo{year}{2012}).

\bibitem{Amo2009}
\bibinfo{author}{Amo, A.} \emph{et~al.}
\newblock \bibinfo{title}{Observation of superfluidity of polaritons in
  semiconductor microcavities}.
\newblock \emph{\bibinfo{journal}{Nat. Phys.}} \textbf{\bibinfo{volume}{5}},
  \bibinfo{pages}{11} (\bibinfo{year}{2009}).

\bibitem{Amo2011}
\bibinfo{author}{Amo, A.} \emph{et~al.}
\newblock \bibinfo{title}{Polariton superfluids reveal quantum hydrodynamic
  solitons.}
\newblock \emph{\bibinfo{journal}{Science}} \textbf{\bibinfo{volume}{332}},
  \bibinfo{pages}{6034} (\bibinfo{year}{2011}).

\bibitem{Carusotto2013}
\bibinfo{author}{Carusotto, I.} \& \bibinfo{author}{Ciuti, C.}
\newblock \bibinfo{title}{Quantum fluids of light}.
\newblock \emph{\bibinfo{journal}{Rev. Mod. Phys.}}
  \textbf{\bibinfo{volume}{85}}, \bibinfo{pages}{299} (\bibinfo{year}{2013}).

\bibitem{Vaupel1996}
\bibinfo{author}{Vaupel, M.}, \bibinfo{author}{Staliunas, K.} \&
  \bibinfo{author}{Weiss, C.~O.}
\newblock \bibinfo{title}{Hydrodynamic phenomena in laser physics: Modes with
  flow and vortices behind an obstacle in an optical channel}.
\newblock \emph{\bibinfo{journal}{Phys. Rev. A}} \textbf{\bibinfo{volume}{80}},
  \bibinfo{pages}{880} (\bibinfo{year}{1996}).

\bibitem{Chiao1999}
\bibinfo{author}{Chiao, R.~Y.} \& \bibinfo{author}{Boyce, J.}
\newblock \bibinfo{title}{Bogoliubov dispersion relation and the possibility of
  superfluidity for weakly interacting photons in a two-dimensional photon
  fluid}.
\newblock \emph{\bibinfo{journal}{Phys. Rev. A}} \textbf{\bibinfo{volume}{60}},
  \bibinfo{pages}{4114} (\bibinfo{year}{1999}).

\bibitem{Pomeau1993}
\bibinfo{author}{Pomeau, Y.} \& \bibinfo{author}{Rica, S.}
\newblock \bibinfo{title}{Diffraction non lin\'eaire}.
\newblock \emph{\bibinfo{journal}{C. R. Acad. Sci. Paris}}
  \textbf{\bibinfo{volume}{397}}, \bibinfo{pages}{1287} (\bibinfo{year}{1993}).

\bibitem{Frisch1992}
\bibinfo{author}{Frisch, T.}, \bibinfo{author}{Pomeau, Y.} \&
  \bibinfo{author}{Rica, S.}
\newblock \bibinfo{title}{Transition to dissipation in a model of superflow}.
\newblock \emph{\bibinfo{journal}{Phys. Rev. Lett.}}
  \textbf{\bibinfo{volume}{69}}, \bibinfo{pages}{11} (\bibinfo{year}{1992}).

\bibitem{Wan2007}
\bibinfo{author}{Wan, W.}, \bibinfo{author}{Jia, S.} \&
  \bibinfo{author}{Fleischer, J.~W.}
\newblock \bibinfo{title}{Dispersive superfluid-like shock waves in nonlinear
  optics}.
\newblock \emph{\bibinfo{journal}{Nat. Phys.}} \textbf{\bibinfo{volume}{3}},
  \bibinfo{pages}{46--51} (\bibinfo{year}{2007}).

\bibitem{Khamis2008}
\bibinfo{author}{Khamis, E.~G.}, \bibinfo{author}{Gammal, A.},
  \bibinfo{author}{El, G.~A.}, \bibinfo{author}{Gladush, Y.~G.} \&
  \bibinfo{author}{Kamchatnov, A.~M.}
\newblock \bibinfo{title}{Nonlinear diffraction of light beams propagating in
  photorefractive media with embedded reflecting wire}.
\newblock \emph{\bibinfo{journal}{Phys. Rev. A}} \textbf{\bibinfo{volume}{78}},
  \bibinfo{pages}{013829} (\bibinfo{year}{2008}).

\bibitem{Leboeuf2010}
\bibinfo{author}{Leboeuf, P.} \& \bibinfo{author}{Moulieras, S.}
\newblock \bibinfo{title}{Superfluid motion of light}.
\newblock \emph{\bibinfo{journal}{Phys. Rev. Lett.}}
  \textbf{\bibinfo{volume}{105}}, \bibinfo{pages}{163904}
  (\bibinfo{year}{2010}).

\bibitem{Carusotto2014}
\bibinfo{author}{Carusotto, I.}
\newblock \bibinfo{title}{Superfluid light in bulk nonlinear media}.
\newblock \emph{\bibinfo{journal}{Proc. R. Soc. A}}
  \textbf{\bibinfo{volume}{470}}, \bibinfo{pages}{0320} (\bibinfo{year}{2014}).

\bibitem{Vocke2015}
\bibinfo{author}{Vocke, D.} \emph{et~al.}
\newblock \bibinfo{title}{Experimental characterization of nonlocal photon
  fluids}.
\newblock \emph{\bibinfo{journal}{Optica}} \textbf{\bibinfo{volume}{2}}
  (\bibinfo{year}{2015}).

\bibitem{Vocke2016}
\bibinfo{author}{Vocke, D.} \emph{et~al.}
\newblock \bibinfo{title}{Role of geometry in the superfluid flow of nonlocal
  photon fluids}.
\newblock \emph{\bibinfo{journal}{Phys. Rev. A}} \textbf{\bibinfo{volume}{94}},
  \bibinfo{pages}{013849} (\bibinfo{year}{2016}).

\bibitem{Leggett1999}
\bibinfo{author}{Leggett, A.}
\newblock \bibinfo{title}{Superfluidity}.
\newblock \emph{\bibinfo{journal}{Rev. Mod. Phys.}}
  \textbf{\bibinfo{volume}{71}}, \bibinfo{pages}{S318 LP}
  (\bibinfo{year}{1999}).

\bibitem{Sun2012}
\bibinfo{author}{Sun, C.} \emph{et~al.}
\newblock \bibinfo{title}{Observation of the kinetic condensation of classical
  waves}.
\newblock \emph{\bibinfo{journal}{Nat. Phys.}} \textbf{\bibinfo{volume}{8}},
  \bibinfo{pages}{470 -- 474} (\bibinfo{year}{2012}).

\bibitem{Allum1977}
\bibinfo{author}{Allum, D.~R.}, \bibinfo{author}{McWlintock, P. V.~E.},
  \bibinfo{author}{Phillips, A.} \& \bibinfo{author}{Bowley, R.~M.}
\newblock \bibinfo{title}{The breakdown of superfluidity in liquid $^{4}$he: an
  experimental test of landau's theory}.
\newblock \emph{\bibinfo{journal}{Phylosophical Transactions of the Royal
  Society of London A}} \textbf{\bibinfo{volume}{284}},
  \bibinfo{pages}{179--224} (\bibinfo{year}{1977}).

\bibitem{Raman1999}
\bibinfo{author}{Raman, C.} \emph{et~al.}
\newblock \bibinfo{title}{Evidence for a critical velocity in a bose-einstein
  condensed gas}.
\newblock \emph{\bibinfo{journal}{Phys. Rev. Lett.}}
  \textbf{\bibinfo{volume}{83}}, \bibinfo{pages}{2502} (\bibinfo{year}{1999}).

\bibitem{Pavloff2002}
\bibinfo{author}{Pavloff, N.}
\newblock \bibinfo{title}{Breakdown of superfluidity of an atom laser past an
  obstacle}.
\newblock \emph{\bibinfo{journal}{Phys. Rev. A}} \textbf{\bibinfo{volume}{66}},
  \bibinfo{pages}{013610} (\bibinfo{year}{2002}).

\bibitem{Miller2007}
\bibinfo{author}{Miller, D.~E.} \emph{et~al.}
\newblock \bibinfo{title}{Critical velocity for superfluid flow across the
  bec-bcs crossover}.
\newblock \emph{\bibinfo{journal}{Phys. Rev. Lett.}}
  \textbf{\bibinfo{volume}{99}}, \bibinfo{pages}{070402}
  (\bibinfo{year}{2007}).

\bibitem{Engels2007}
\bibinfo{author}{Engels, P.} \& \bibinfo{author}{Atherton, C.}
\newblock \bibinfo{title}{Stationary and nonstationary fluid flow of a
  bose-einstein condensate through a penetrable barrier}.
\newblock \emph{\bibinfo{journal}{Phys. Rev. Lett.}}
  \textbf{\bibinfo{volume}{99}}, \bibinfo{pages}{160405}
  (\bibinfo{year}{2007}).

\bibitem{Desbuquois2012}
\bibinfo{author}{Desbuquois, R.} \emph{et~al.}
\newblock \bibinfo{title}{Superfluid behaviour of a two-dimensional bose gas}.
\newblock \emph{\bibinfo{journal}{Nat. Phys.}} \textbf{\bibinfo{volume}{8}},
  \bibinfo{pages}{645--648} (\bibinfo{year}{2012}).

\bibitem{Wouters2010}
\bibinfo{author}{Wouters, M.} \& \bibinfo{author}{Carusotto, I.}
\newblock \bibinfo{title}{Superfluidity and critical velocities in
  nonequilibrium bose-einstein condensates}.
\newblock \emph{\bibinfo{journal}{Phys. Rev. Lett.}}
  \textbf{\bibinfo{volume}{105}}, \bibinfo{pages}{020602}
  (\bibinfo{year}{2010}).

\bibitem{Berceanu2012}
\bibinfo{author}{Berceanu, A.~C.}, \bibinfo{author}{Cancellieri, E.} \&
  \bibinfo{author}{Marchetti, F.~M.}
\newblock \bibinfo{title}{Drag in a resonantly driven polariton fluid}.
\newblock \emph{\bibinfo{journal}{J. Phys.: Condens. Matter}}
  \textbf{\bibinfo{volume}{24}}, \bibinfo{pages}{235802}
  (\bibinfo{year}{2012}).

\bibitem{VanRegemortel2014}
\bibinfo{author}{Van~Regemortel, M.} \& \bibinfo{author}{Wouters, M.}
\newblock \bibinfo{title}{Negative drag in nonequilibrium polariton quantum
  fluids}.
\newblock \emph{\bibinfo{journal}{Phys. Rev. B}} \textbf{\bibinfo{volume}{89}},
  \bibinfo{pages}{085303} (\bibinfo{year}{2014}).

\bibitem{Larre2012}
\bibinfo{author}{Larr{\'{e}}, P.-{\'{E}}.}, \bibinfo{author}{Pavloff, N.} \&
  \bibinfo{author}{Kamchatnov, A.}
\newblock \bibinfo{title}{Wave pattern induced by a localized obstacle in the
  flow of a one-dimensional polariton condensate}.
\newblock \emph{\bibinfo{journal}{Phys. Rev. B}} \textbf{\bibinfo{volume}{86}},
  \bibinfo{pages}{165304} (\bibinfo{year}{2012}).

\bibitem{Larre2017}
\bibinfo{author}{Larr{\'e}, P.~E.}, \bibinfo{author}{Biasi, S.},
  \bibinfo{author}{Ramiro-Manzano, F.}, \bibinfo{author}{Pavesi, L.} \&
  \bibinfo{author}{Carusotto, I.}
\newblock \bibinfo{title}{Pump-and-probe optical transmission phase shift as a
  quantitative probe of the bogoliubov dispersion relation in a nonlinear
  channel waveguide}.
\newblock \emph{\bibinfo{journal}{Eur. Phys. J. D}}
  \textbf{\bibinfo{volume}{71}}, \bibinfo{pages}{146} (\bibinfo{year}{2017}).

\bibitem{Feynman1955}
\bibinfo{author}{Feynman, R.}
\newblock \emph{\bibinfo{title}{Progress in Low Temperature Physics}},
  vol.~\bibinfo{volume}{1} (\bibinfo{publisher}{North-Holland, Amsterdam},
  \bibinfo{year}{1955}).

\bibitem{Carusotto2006}
\bibinfo{author}{Carusotto, I.}, \bibinfo{author}{Hu, S.~X.},
  \bibinfo{author}{Collins, L.~A.} \& \bibinfo{author}{Smerzi, A.}
\newblock \bibinfo{title}{Bogoliubov-cerenkov radiation in a bose-einstein
  condensate flowing against an obstacle}.
\newblock \emph{\bibinfo{journal}{Phys. Rev. Lett.}}
  \textbf{\bibinfo{volume}{97}}, \bibinfo{pages}{260403}
  (\bibinfo{year}{2006}).

\bibitem{Larre2015}
\bibinfo{author}{Larr{\'{e}}, P.-{\'{E}}.} \& \bibinfo{author}{Carusotto, I.}
\newblock \bibinfo{title}{Optomechanical signature of a frictionless flow of
  superfluid light}.
\newblock \emph{\bibinfo{journal}{Phys. Rev. A}} \textbf{\bibinfo{volume}{91}},
  \bibinfo{pages}{053809} (\bibinfo{year}{2015}).

\bibitem{albert2008}
\bibinfo{author}{Albert, M.}, \bibinfo{author}{Paul, T.},
  \bibinfo{author}{Pavloff, N.} \& \bibinfo{author}{Leboeuf, P.}
\newblock \bibinfo{title}{Dipole oscillations of a bose-einstein condensate in
  the presence of defects and disorder}.
\newblock \emph{\bibinfo{journal}{Phys. Rev. Lett.}}
  \textbf{\bibinfo{volume}{100}}, \bibinfo{pages}{250405}
  (\bibinfo{year}{2008}).

\bibitem{Buse1997}
\bibinfo{author}{Buse, K.}
\newblock \bibinfo{title}{Light-induced charge transport processes in
  photorefractive crystals ii: Materials}.
\newblock \emph{\bibinfo{journal}{Appl. Phys. B}}
  \textbf{\bibinfo{volume}{64}}, \bibinfo{pages}{391--407}
  (\bibinfo{year}{1997}).

\bibitem{Denz2003}
\bibinfo{author}{Denz, C.}, \bibinfo{author}{Schwab, M.} \&
  \bibinfo{author}{Weilnau, C.}
\newblock \emph{\bibinfo{title}{Transverse-Pattern Formation in Photorefractive
  Optics}} (\bibinfo{publisher}{Springer-Verlag, Berlin},
  \bibinfo{year}{2003}).

\bibitem{Carusotto2013b}
\bibinfo{author}{Carusotto, I.} \& \bibinfo{author}{Rousseaux, G.}
\newblock \emph{\bibinfo{title}{The C̆erenkov Effect Revisited: From Swimming
  Ducks to Zero Modes in Gravitational Analogues}}, \bibinfo{pages}{109 -- 144}
  (\bibinfo{publisher}{Springer International Publishing},
  \bibinfo{year}{2013}).

\end{thebibliography}

\vspace{1cm}
\noindent{\bf Aknowledgments}
The paper is dedicated to the memory of Patricio Leboeuf who was very enthusiastic about the idea of superfluid motion of light. M. A., will always be grateful to him for his kindness, trust and freedom he gave to him when he was his PhD student at LPTMS. 
The authors acknowledge helpful contributions from M. Garsi during the early stage of this work. We also thank I. Carusotto, V. Doya, F. Mortessagne, N. Pavloff and P. Vignolo for helpful discussions. This work has been supported by the the Region PACA and the French government, through the UCA$^\text{{\tiny JEDI}}$ Investments in the Future project managed by the National Research Agency (ANR) with the reference number ANR-15-IDEX-01. P.-\'E.~L. was funded by the Centre National de la Recherche Scientifique (CNRS) and by the ANR under the grant ANR-14-CE26-0032 LOVE.

\bigskip
\noindent{\bf Author contributions}
C.M., O.B. and M.B. performed the experiments and analyzed the data. M.A. and P.\'E.L. developed the theory. All authors participated in the discussions and in writing the paper.

\bigskip
\noindent{\bf Correspondence} Correspondence and requests for materials should be addressed to C.M. (email: claire.michel@unice.fr) and M.B. (email: bellec@unice.fr).
 
\clearpage

\appendix

\makeatletter
\renewcommand{\thesection}{S\arabic{section}}
\renewcommand{\thesubsection}{\thesection.\arabic{subsection}}
\renewcommand{\thefigure}{S\arabic{figure}}
\renewcommand{\theequation}{S.\arabic{equation}}
\makeatother

\setcounter{figure}{0}
\setcounter{equation}{0}

\section*{SUPPLEMENTARY INFORMATION}

\noindent \textbf{S1. Expression of the fluid velocity}
\medskip

In Eq.~(\ref{eq:schro}) of the main text, the gradient of the phase of the complex envelope $\Ef$ of the electric field $\mathrm{Re}(\Ef\,e^{i\nr\kf z})$ play the role of the velocity $v=(\nr\,\kf)^{-1}\,|\partial\arg(\Ef)/\partial\mathbf{r}|$ of the fluid of light.
In the experiment, the fluid-of-light beam consists in a gaussian beam which is large compared to the size of the obstacle (see Fig.~\ref{fig:sketch}a). It can be approximated by a plane wave, such as $\Ef\propto e^{i\mathbf{k_\perp}\cdot\mathbf{r}}$, with $k_\perp=\kf\sin{\theta_{\rm in}}$ the transverse wave vector of the plane wave. Its phase $\mathbf{k_\perp}\cdot\mathbf{r}$ thus remains constant in the vicinity of the obstacle. Consequently, $v$ is only given by $\theta_{\rm in}$, the angle between the fluid-of-light beam and the $z$ direction: $v=\sin{\theta_{\rm in}}/\nr\simeq\theta_{\rm in}/\nr$ in the here-considered paraxial approximation.

\vspace{1em}
\noindent \textbf{S2. Dispersion relation and healing length}
\medskip

\begin{figure}[b]
	\centering
	\includegraphics[scale=1]{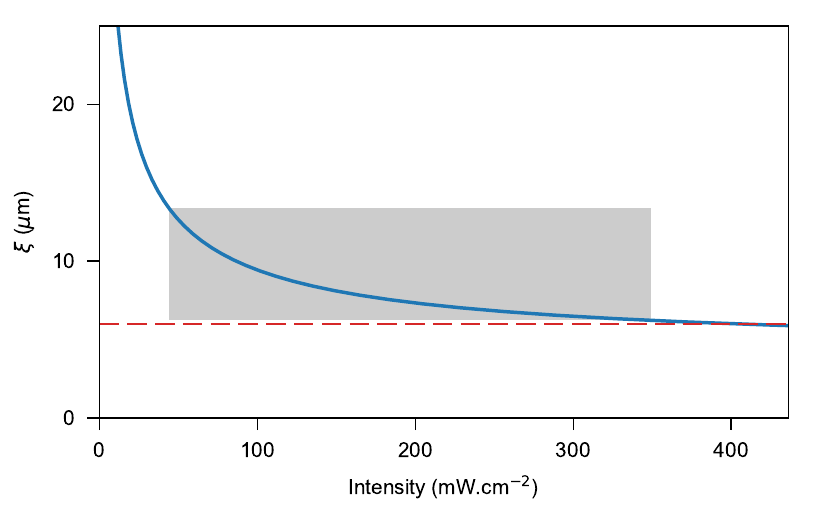}
	\caption{\textbf{Theoretical variation of the healing length $\xi$ with respect to the intensity}. The dashed red line corresponds to the radius of the obstacle. The gray box defines the experimentally accessible range.}
	\label{fig_supp:xi}
\end{figure}

Considering the Bogoliubov theory of weak perturbations on top of a uniform fluid of light, the dispersion relation reads \cite{Carusotto2013b, Carusotto2014}
\begin{equation}
	\mathcal{W}(k_\perp)=\sqrt{\frac{k_\perp^2}{2\nr \kf}\left( \frac{k_\perp^2}{2\nr \kf} + \kf \Delta n(\If) \right)}
\end{equation}
The quantity $\xi=[\nr\,\kf\times\kf\,|\Delta n(\If)|]^{-1/2}$ is called the healing length and can thus be extracted from the previous equation. It defines a length scale for the smallest intensity modulations that can occur in the system.
In the main text, the size of the obstacle is compared to the healing length of the fluid of light. Nevertheless, this quantity is intensity-dependent in the here-considered case. The calculated values are plotted in Fig.~\ref{fig_supp:xi}. The red dashed line corresponds to the radius of the obstacle (estimated experimentally at 6 $\mu$m).

\vspace{1em}
\noindent \textbf{S3. Optical analog of the Ehrenfest relations}
\medskip

The Ehrenfest theorem relates the time derivative of the expectation values of the position and momentum operators $x$ and $p$ to the expectation value of the force $F = - dV/dx$ on a massive particle moving in a scalar potential,
\begin{equation}
	m\frac{d}{dt}\langle x \rangle=\langle p \rangle,\;\frac{d}{dt}\langle p \rangle = -\left\langle \frac{\partial V(x)}{\partial x}\right\rangle
\end{equation}
while the equation describing the motion of a massive particle in a potential is the Schr\"odinger equation $i\hbar\partial_t\psi=-\frac{\hbar^2}{2m}\Delta\psi+V(x,t)\psi$ and the Hamiltonian reads $H(x,p,t)=\frac{p^2}{2m}+V(x,t)$.
\begin{figure}[b]
	\centering
	\includegraphics[scale=1.4]{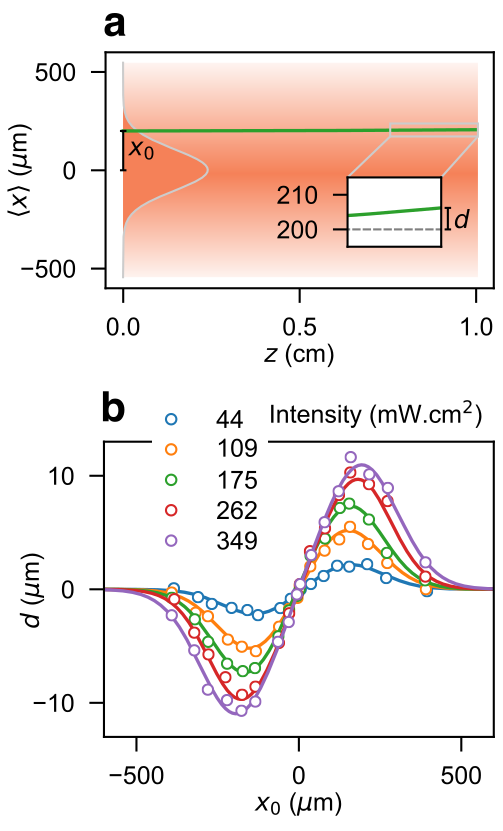}
	\caption{\textbf{Transverse displacement of the obstacle in a gaussian potential.} 
	(\textbf{a}) Calculated transverse displacement $\langle x \rangle$ along the propagation, $z$, axis for a potential induced by a 270 $\mu$m at 1/$e^2$ half-width gaussian laser beam of intensity $\If = 175$ mW.cm$^{-2}$. $x_0 = 200$ $\mu$m is the initial position of the obstacle. $d$ is the transverse displacement, with respect to $x_0$, at the crystal's output. 
	(\textbf{b}) Measured transverse displacement for various laser beam intensities $\If$ ranging from $44$ to $349$ mW.cm$^{-2}$ as a function of $x_0$. The fit procedure (solid lines) allows to extracted $\Is = 380 \pm 50$ W.cm$^{-2}$ and $\Delta n_{\rm max} = 2.5 \pm 0.4 \ttt{-4}$.}
	\label{fig_supp:displacement}
\end{figure}
Using the analogy between the Schr\"odinger equation and the linear propagation equation for the obstacle beam in the optical potential $\Delta n(\If)$ which reads
\begin{equation}
	i\partial_z \Ed = -\frac{1}{2\nr\kd}\nabla^2 \Ed - \kd \Delta n(\If) \Ed
	\label{eq:schro_defect}
\end{equation}
we can write the Hamiltonian which takes the form
\begin{equation}
	H(x,k_\perp,z)=\frac{k_\perp^2}{2\nr\kd}-\kd\Delta n(\If)
	\label{eq:ham}
\end{equation}
Denoting $\langle x \rangle = \int x |\Ed|^2 \, dx$ the position of the centroid of the obstacle beam, the optical analog of the Ehrenfest theorem reads:
\begin{eqnarray}
	& &\left(\nr\kd\right)\frac{d}{dz}\langle x \rangle = \langle k_\perp \rangle\\
	& &\frac{d}{dz}\langle k_\perp \rangle = -\left\langle -\kd\frac{\partial\Delta n(\If)}{\partial x}\right\rangle
\end{eqnarray}
which leads to the equation of motion for the centroid of the obstacle beam:
\begin{equation}
(\nr\,\kd)\,\partial_{zz}\langle x \rangle = \partial_x [\kd\,\Delta n(\If)]
\end{equation}
{\em Obstacle displacement in the fluid of light at rest.}---
By assuming that $\Delta n$ is $z$ independent, which is valid in the here-considered linear propagation of the obstacle beam, we readily obtain
\begin{equation}
	d=\langle x(z) \rangle - x_0 = \frac{1}{2}\,[\partial_x \Delta n(\If)/\nr]\,z^2
	\label{eq:displacement}
\end{equation}
where $x_0$ is the initial position of the obstacle. 

\noindent As shown in Fig.~\ref{fig_supp:displacement}a, for $x_0 = 200$ $\mu$m and an optical potential induced by a 270 $\mu$m wide gaussian beam of intensity $\If = 175$ mW.cm$^{-2}$, the relative transverse displacement $d$ defined in \refeq{eq:displacement} reaches 7.8 $\mu$m at the output of the crystal (see inset).
The experimental measurement of $d$, for various intensities and positions $x_0$, is presented in Fig.~\ref{fig_supp:displacement}b. The experimental data are fitted (solid lines), using the above expression, the saturation intensity and the maximum refractive index modification being the fitting parameters. We extract $\Is = 380 \pm 50$ mW.cm$^{-2}$ and $\Delta n_{\rm max} = 2.5 \pm 0.4 \ttt{-4}$. 
It is worth mentioning that the value of $\Is$ is used for the calculation of $\Delta n(I) $ and its deriving quantities (i.e., $\cs$ and $\xi$).\\

\begin{figure}[t]
	\centering
	\includegraphics[scale=1.6]{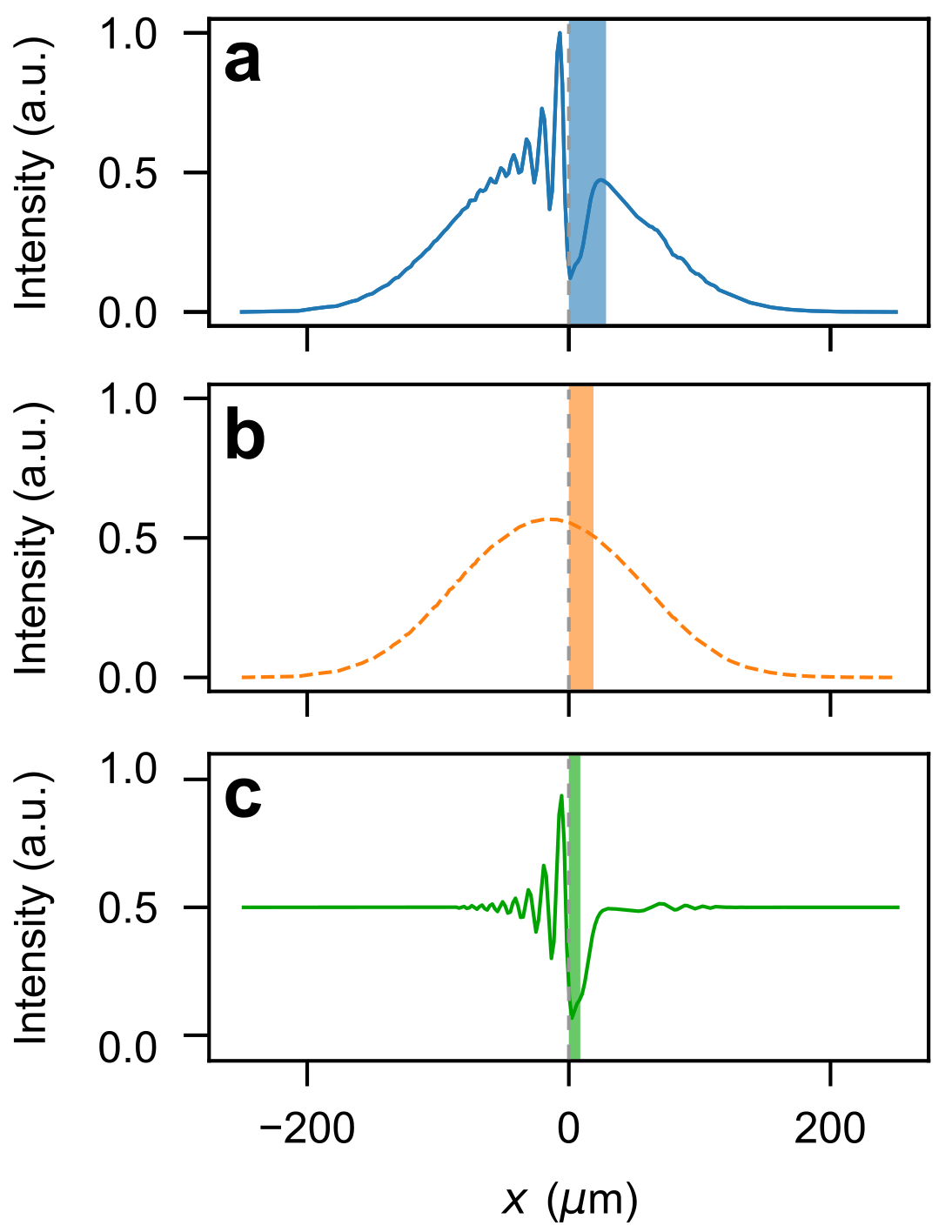}
	\caption{\textbf{Illustration of the measurement of the net obstacle displacement.} In each panel, the profile corresponds to a cut along the $x$ axis of the spatial distribution of intensity. The dashed grey line corresponds to the all-optical obstacle position. 
	(\textbf{a}) Typical profile of the fluid of light intensity. The total corresponding transverse displacement is illustrated by the blue region.  
	(\textbf{b}) Typical profile of the unperturbed fluid of light intensity (i.e. when the obstacle is weak). The transverse displacement induced by the gaussian shape of the fluid of light is depicted by the orange region.
	(\textbf{c}) The net transverse displacement of the obstacle induced by the local intensity modulation (green curve) is shown by the green region.
	}
	\label{fig_supp:disp_corr}
\end{figure}

\noindent{\em Obstacle displacement in a moving fluid of light.}---
In Fig.~\ref{fig:displacement_local_int}.b of the main text, the net transverse displacement of the obstacle induced by the local modulation of the fluid of light intensity is plotted for various initial conditions. As seen previously, the all-optical obstacle is highly sensitive to the surrounding refractive index potential resulting from the gaussian distribution of the fluid of light intensity. Thus, for each data point, we subtract the displacement measured at very low obstacle intensity, when its influence on the fluid of light is negligible. This correction procedure is illustrated in Fig.~\ref{fig_supp:disp_corr}.
Figure~\ref{fig_supp:disp_corr}a represents a cut along the $x$-axis of a spatial distribution of intensity in the supersonic case. The blue box reporesents the displacement of the obstacle between the input and the output of the crystal. Figure~\ref{fig_supp:disp_corr}b shows the displacement of the obstacle for the same initial condition but without any influence of the obstacle on the fluid of light. Figure~\ref{fig_supp:disp_corr}c depicts the net displacement obtained from Fig.~\ref{fig_supp:disp_corr}a and corrected by subtracting the intrinsic effect of the gaussian shape of the fluid-of-light beam (Fig.~\ref{fig_supp:disp_corr}b).

\vspace{1em}
\noindent \textbf{S4. Qualitative discussion on the obstacle extra-displacement}
\medskip

\begin{figure*}
	\centering
	\includegraphics[scale=1.4]{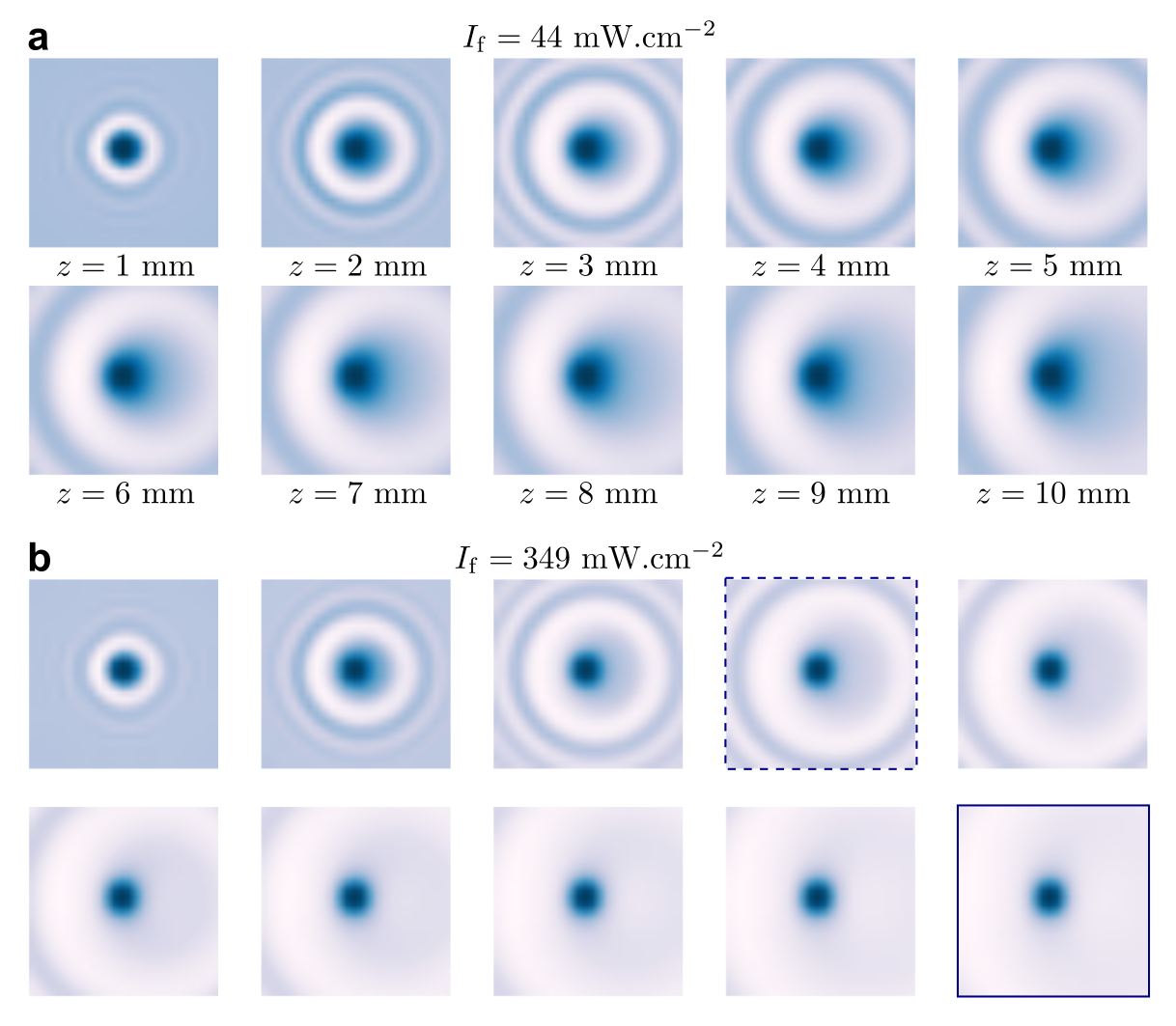}
	\caption{\textbf{Numerical simulations of the evolution of the fluid of light}. In each panel, the images correspond to the snapshots of the fluid of light intensity taken at various distances $z$ ranging from 1 to 10 mm for two different input intensities. (a) $\If = 44$ mW.cm$^{-2}$ and (b) $\If = 349$ mW.cm$^{-2}$. The input angle is fixed at $\theta_{\rm in}=5$ mrad. The corresponding Mach numbers are respectively 0.7 and 0.3. The image size is 50$\times$50 $\mu$m$^2$. The center of the images correspond to the position of the obstacle.}
	\label{fig_supp:sim}
\end{figure*}

In Fig.~\ref{fig:displacement_local_int}b of the main text, we observe that the displacement is not purely zero in the superfluid regime for large intensities. We claim that this is likely due to the displacement acquired during the non-stationary regime at the early stage of the propagation.
Here, by means of numerical simulations giving access to the evolution along $z$ of the fluid of light field, we qualitatively discuss this argument.
The images in Fig.~\ref{fig_supp:sim} shows numerical simulation of the evolution of the fluid of light obtained for $\theta_{\rm in}=5$ mrad and (a) $\If = 44$ mW.cm$^{-2}$, i.e. $v/\cs = 0.7$ and (b) $\If = 349$ mW.cm$^{-2}$, i.e. $v/\cs = 0.3$. Each panel corresponds to the snapshots of the fluid of light intensity taken at various distances $z$ ranging from 1 to 10 mm. The image size is reduced to 50$\times$50 $\mu$m$^2$ to get focused on the intensity distribution in the vicinity of the obstacle.
At large intensity, according to the symmetric intensity distribution observed for $z=10$ mm (blue line panel Fig.~\ref{fig_supp:sim}b), no displacement is expected. 
However, the transient regime shows an asymmetric intensity distribution in the vicinity of the obstacle (dashed line panel Fig.~\ref{fig_supp:sim}b). The associated refractive index modification may lead to transverse displacement of the obstacle. 

\end{document}